\documentclass[11pt]{article}

\usepackage{geometry}                
\usepackage{graphicx}
\usepackage{amsmath,amssymb, amsthm,epsfig,bm}
\usepackage{setspace}
\usepackage{epstopdf}
\usepackage{psfrag}
\usepackage{color,soul}
\usepackage{verbatim}
\usepackage{multirow}
\usepackage{natbib}
\usepackage{xr}
\usepackage{mathrsfs}
\usepackage{cases}
\usepackage{enumitem}

\setlist{noitemsep, topsep=0pt}
\bibpunct{(}{)}{;}{a}{}{,} 
\usepackage[english]{babel}
\usepackage[utf8]{inputenc}
\usepackage{fancyhdr}
 
\newcommand{\titleshort}{Dependence Thresholding for Nonlinear Latent Factor Models}
\newcommand{\authorshort}{Kim and Zhou}

\usepackage[hidelinks, unicode]{hyperref} 
\usepackage{multirow}

\numberwithin{equation}{section}

\newtheorem{theorem}{Theorem}
\newtheorem{proposition}{Proposition}
\newtheorem{lemma}{Lemma}

\theoremstyle{definition}
\newtheorem{definition}{Definition}
\newtheorem{remark}{Remark}
\newtheorem{condition}{Condition}
\newtheorem{example}{Example}

\DeclareMathOperator{\ch}{ch}
\DeclareMathOperator{\pa}{pa}
\DeclareMathOperator{\Var}{Var}

\usepackage{bm} 
\usepackage{graphicx} 
\usepackage{multirow} 
\usepackage[linesnumbered, vlined, ruled]{algorithm2e}
\usepackage{amssymb} 
\usepackage[colorinlistoftodos, textwidth = 20mm]{todonotes} 
\usepackage{subcaption} 
\usepackage{enumitem} 
\usepackage{amsthm}
\usepackage{appendix}
\usepackage{placeins} 
\usepackage{mathtools}
\usepackage{siunitx} 
\usepackage{booktabs}

\newcommand{\E}{\mathbb{E}}
\newcommand{\Prob}{\mathbb{P}}
\newcommand{\R}{\mathbb{R}}
\newcommand{\indep}{\mathrel{\perp\!\!\!\perp}}

\begin{document}

\title{Identification of Nonlinear and Dependent Latent Factor Structure through Clique Search}
\author{%
\begin{tabular}[t]{c}
  Dale S.~Kim\thanks{UCLA Department of Statistics \& Data Science. Email: daleskim@stat.ucla.edu; zhou@stat.ucla.edu}
  \quad and \quad
  Qing Zhou$^*$\\
\end{tabular}%
}
\date{}
\maketitle

\begin{abstract}
Learning the structure of latent factor models involves two central challenges: (1) estimating the number of latent factors and (2) learning the support of the mapping from latent variables to observed variables.
This is especially challenging for nonparametric regimes and nonlinear settings.
We propose a method for latent structure learning, based on pairwise dependence measures on the observed variables using a graph-theoretic representation.
We show that both the number of latent factors and nonlinear mapping structure can be identified from the distribution of observed variables under mild structural assumptions.
Unlike prior work restricted to linear correlations, we establish identifiability and consistency for a general class of dependence measures under nonlinear factor models.
This motivates a Dependence Thresholding (DT) algorithm, which jointly estimates the number of latent factors and nonlinear mapping structure from observational data alone.
We pair this with a neural network architecture constrained by the nonlinear mapping structure, to recover the nonlinear function.
Through simulation studies, we show that the DT algorithm is accurate in practice, even when using flexible methods such as neural networks, and exhibits robustness against violations of its assumptions in high-dimensional settings.

\end{abstract}

\section{Introduction} \label{sec:introduction}

Understanding the underlying latent mechanisms that generate high-dimensional observations is a central goal of scientific inquiry and statistical modeling.
Much work has been dedicated to recovering latent representations across various settings.
In the context of nonlinear independent components analysis, latent representations can be recovered under various sparsity conditions and assumptions on the nonlinear mixing function \citep{Zheng2022, Zheng2023}.
Alternatively, auxiliary variables can be utilized to identify the latent variables \citep{Hyvarinen2016, Khemakhem2020}.
However, these methods require the number of latent factors to be known, a problem that has received less attention.
From the perspective of causal representation learning \citep{Scholkopf2022, Zhang2024, Ahuja2023}, the problem of learning the dimensionality of latent variables is of fundamental importance, as well as the functional relationships between the latent and observed variables.

In this work, we focus on learning the structure of the latent factor model, which entails both the number of factors and the relations between the latent and observed variables, where we will call the latter the mapping structure.
In the linear case, estimating the number of latent variables has classically been done through spectral analysis of the covariance or correlation matrices of the observed variables \citep{Kaiser1960, Horn1965, Cattell1966, Guttman1954} or model selection methods \citep{Preacher2013, Bartlett1950}.
For the mapping structure, a common practice is to first estimate a saturated model, then threshold low coefficients to zero \citep{Howard2016,Ford1986}.
In a similar spirit, more modern methods optimize a penalized objective to learn the structure using various penalty functions \citep{Trendafilov2017, Hirose2014, Hirose2022}.
However, one drawback to these methods is that the number of latent variables and the mapping structure are handled separately, all with varying degrees of criticisms \citep{Auerswald2019, Browne1968, Scharf2019}, and they are restricted to linear factor models.

More recently, a few works have emerged utilizing graph-theoretic representations to learn the number of factors and the mapping structure simultaneously, with possibly nonlinear generalizations.
For independent factors, \cite{Markham2020} uses clique covers to learn a lower-bound on the number of factors and the accompanying mapping structure.
Allowing factors to be causally related, \cite{Jiang2023} uses cliques observed in interventional data to learn the structure of the latent factor model and an equivalence class of causal models on the latent variables.
Using only observational data and allowing factors to be correlated, \cite{Kim2023} utilize a notion of uniquely maximal cliques to learn the latent factor structure, but rely on linear mapping functions.

In this work, we focus on the case of (1) allowing correlated factors, (2) not requiring interventional or auxiliary information, and (3) allowing arbitrary nonlinear mapping functions.
Let $X = (X_1, \dots, X_p) \in \R^{p}$ be a vector of observed variables, and $Z = (Z_1, \dots, Z_d) \in \R^{d}$ be a vector of latent variables. 
We consider a general latent factor model:
\begin{equation} \label{model}
X = g(Z, \varepsilon),
\end{equation}
where $\varepsilon = (\varepsilon_1, \dots, \varepsilon_p)$ is a vector of independent errors with diagonal $\Var(\varepsilon) = \Omega$.
The mapping $g = (g_1, \dots, g_p)$ is in general nonlinear, $g_i$ is a function of $(Z, \varepsilon_i)$, and $g_i$ admits first-order partial derivatives with respect to the latent variables.
We denote the covariance matrix of the latent variables as $\Var(Z) = \Phi$ and we assume that $d < p$.
It is assumed throughout that latent indices are arbitrary, and can be permuted without the loss of generality.
We construct graphs induced by general pairwise dependence measures among $X_1,\ldots,X_p$, and show that the support of the nonlinear map $g$ is in one-to-one correspondence with a special class of maximal cliques in these graphs under mild structural assumptions.
This correspondence reduces the problem of learning the mapping structure to a simple and computationally efficient clique search problem.
Once the structure is recovered, it can then be used to constrain the architectures of flexible estimators, such as neural networks, to learn the nonlinear function $g$.

Note that prior work in \cite{Kim2023} is a special case of our learning procedure, in which $g$ is a linear mixing of $Z$ with additive Gaussian errors $\varepsilon$.
The current research presents a substantial generalization of their work.
Unlike the linear Gaussian setting, nonlinear latent factor models generally do not admit tractable observed-data likelihoods or covariance representations. Consequently, both estimation and model selection become substantially more challenging, requiring flexible nonlinear estimators and latent variable inference.
Moreover, structure recovery can no longer be handled through correlation thresholded graphs, motivating our use of general dependence-based graphs and accompanying theoretical development.

Our contributions are as follows:
\begin{enumerate}
  \item Provide the sufficient conditions under which a general latent factor structure is identifiable from general pairwise dependence measures.
  \item Establish high-dimensional statistical guarantees of recovering the latent factor structure under a nonlinear mapping.
  \item Propose an algorithm under which the latent factor structure can be estimated from observational data alone.
\end{enumerate}
The remainder of the paper is as follows.
In Section~\ref{sec:preliminaries}, we formalize the latent-to-observed mapping structure and the preliminary graph-theoretic notation and objects used to learn the nonlinear factor model.
In Section~\ref{sec:structure_identifiability}, we establish the correspondence between our structure and the so-called parent-sharing graph used in our method.
In Section~\ref{sec:stats}, we study the statistical recovery of this graph and the structure of the model using general pairwise dependence measures, and derive finite-sample error bounds and high-dimensional consistency results.
In Section~\ref{sec:algorithm}, we present the Dependence Thresholding (DT) algorithm and present methodology to perform structure-constrained nonlinear estimation and model selection.
Finally, in Section~\ref{sec:simulation}, we evaluate the proposed framework through simulation studies, and conclude with a discussion in Section~\ref{sec:discussion}.
All proofs may be found in the appendix.

\section{Preliminaries} \label{sec:preliminaries}

We are interested in learning which latent variables affect each observed variable.
That is, we are interested in the support of the Jacobian matrix $J_g(z)$.
Formally, we define this support as follows.  
\begin{definition}[Jacobian Support]
\label{def:jacobian_support}
\begin{equation}
A_0 \coloneqq \left\{ (i, j) : \exists\, z \text{ such that } \dfrac{\partial g_i(z)}{\partial z_j} \neq 0 \right\}.
\end{equation}
Without loss of generality, we assume there are no unused latent variables.
That is, for each $j \in \{1, \dots, d\}$, there exists an $i \in \{1, \dots, p\}$ such that $(i, j) \in A_0$.
\end{definition}

If the $(i, j)$ entry of $J_g(z)$ is supported, then we will say that $Z_j$ is a \textit{parent} of $X_i$ and $X_i$ is a \textit{child} of $Z_j$.
We define these as their own sets:
\begin{equation}\label{eq:pach}
\begin{aligned}
\pa(X_i) &\coloneqq \left\{Z_j : (i, j) \in A_0 \right\}\\
\ch(Z_j) &\coloneqq \left\{X_i : (i, j) \in A_0 \right\}.
\end{aligned}
\end{equation}

To learn the support of $J_g(z)$, it will be useful to represent the $\ch(Z_j)$ sets as a structure in a graph that is constructed from $X$ alone.
We use the following graph theoretic notation.
We define a graph $\mathcal{G}$ as an ordered pair $(V, E)$, explicitly denoted as $\mathcal{G}(V, E)$, where $V$ is a set of vertices and $E \subseteq V \times V$ is a set of edges.
For convenience, we will use $V = X$ to mean that the elements of the vertex set $V$ represent the index set of the random vector $X$.
We also restrict our attention to undirected graphs.
A \textit{clique} of $\mathcal{G}(V, E)$ is a subset of vertices $C \subseteq V$ such that all pairs of distinct vertices in $C$ are connected by an edge.
A \textit{maximal clique} is a clique that cannot be extended by including more vertices from $V$.
Additionally, we will utilize a special kind of maximal clique which we term as \textit{independent maximal clique}:
\begin{definition}[Independent Maximal Clique]
\label{def:independent_maximal_clique}
Let $\mathcal{C} = \{C_1, \dots, C_k\}$ be the set of all maximal cliques in a graph $\mathcal{G}$.
Then, $C_i$ is an independent maximal clique if
\begin{equation}
C_i \nsubseteq \bigcup_{j \neq i} C_j.
\end{equation}
\end{definition}
Essentially, an independent maximal clique is a maximal clique that contains a vertex that is not a member of any other maximal clique.
We call such a vertex a \textit{unique member} of the independent maximal clique.
We use the word ``independent'' as an analog to the notion of linear independence in a vector space.
That is, an independent maximal clique cannot be covered by the union of any of the other maximal cliques.

\section{Structural Identifiability via Parent-Sharing Graphs} \label{sec:structure_identifiability}

We begin by studying relationships among the observed variables that are implied by the latent factor model.
We do this by categorizing pairs of observed variables as sharing parents, or not sharing parents, by constructing a so-called parent-sharing graph.
Using a similar line of reasoning to \cite{Kim2023}, we show that under suitable structural assumptions, the support of the Jacobian is in one-to-one correspondence with the independent maximal cliques of the parent-sharing graph.

We establish the relationship between the structure of the latent factor model ($A_0$) and a \textit{parent-sharing} graph $\mathcal{G}_0(X, E_0)$ defined as follows.
\begin{definition}[Parent-Sharing Graph]
The parent-sharing graph is defined as $\mathcal{G}_0(X, E_0)$, where
\begin{equation}
E_0 \coloneqq \{ (i, j) : \mathrm{pa}(X_i) \cap \mathrm{pa}(X_j) \neq \emptyset \}.
\end{equation}
\end{definition}
Thus, $E_0$ simply encodes whether two observed variables share at least one latent parent.

We note that without further assumptions, multiple latent structures induce the same parent-sharing graph.
We rule out such ambiguities with a sparsity condition called the \textit{unique child condition}:
\begin{condition}[Unique Child Condition] \label{con:ucc}
A latent factor model is said to follow the unique child condition if for all $Z_k$ there exists a $U_k \in X$ such that
\begin{equation} \label{eq:uc_def}
U_k \coloneqq \ch(Z_k) - \bigcup_{j \neq k} \ch(Z_j) \neq \emptyset,\quad\quad \forall\; k\in[d],
\end{equation}
where $U_k$ are called the unique children for $Z_k$.
\end{condition}

This condition yields a one-to-one correspondence between the latent structure and the parent-sharing graph.
We state this formally with the following lemma.
\begin{lemma}[Structural Identifiability via Parent-Sharing Graphs]
\label{lem:ucc_imc_bijection}
If the unique child condition holds, then the set of child sets $\{\ch(Z_j):j \in [d]\}$ is identical to the set of independent maximal cliques in $\mathcal{G}_0$.
\end{lemma}

This result shows that the latent factor model can be obtained from the parent-sharing graph on $X$.
Let $\mathcal{C}(E) = \{C_1, \dots, C_d\}$ denote the set of independent maximal cliques in the graph $\mathcal{G}(X, E)$.
Lemma~\ref{lem:ucc_imc_bijection} shows that the set $\mathcal{C}(E)$ is identical to the set $\{\ch(Z_j):j \in [d]\}$.
Thus, we may reconstruct the Jacobian support by setting
$A = \{(i, j) : i \in \ch(Z_j)\}$.
We may shorthand this procedure as a map $\psi: E\mapsto A$.

We briefly remark that previous works have required exactly one unique child \citep{Saeed2020}, two unique children \citep{Moran2022, Bing2020,Cai2019}, or more \citep{Shimizu2009, Kummerfeld2016} to learn the number of latent variables.
In contrast, our unique child condition is more relaxed, requiring only that each latent factor has at least one unique child.
The usage of unique children arises frequently in psychological and educational measurement settings, where survey instruments are designed such that items are only affected by a single latent construct \citep{Hattie1985, Anderson1988}.
Further, in biological applications, gene-expression measurements are used to uniquely define latent biological pathways \citep[e.g.,][]{Carvalho2008}.
We may also consider a probabilistic graph setting, letting $A$ represent a set of bipartite edges between $Z$ and $X$.
If the parents are randomly assigned to children, then as long as there are sufficiently more observed variables than latent variables, each factor is very likely to have at least one unique child.
This can be seen in the following proposition.

\begin{proposition}[Unique Child Condition under Random Parent Assignment]
\label{prop:ucc_random}
Consider a bipartite graph between vertex sets $Z$ and $X$.
Denote the probability of an edge between each $X_i$ and each $Z_j$ as $\alpha \in (0, 1)$, all mutually independent.
If every $X_i$ has at least one parent, then
\begin{equation}
\label{eq:ucc_random_bound}
\Prob(\mathrm{UCC}) \geq 1 - d\exp\bigl(-p\alpha(1-\alpha)^{d}\bigr),
\end{equation}
where $\mathrm{UCC}$ denotes Condition~\ref{con:ucc} holding.
Additionally, if the graph is sparse such that $\alpha = c/d$ with a constant $c > 0$, then
\begin{equation} \label{eq:ucc_random_sparse}
\Prob(\mathrm{UCC}) \geq 1 - d\exp\left(-c_1 p/d\right),
\end{equation}
for sufficiently large $d$, where $c_1 = c \exp(-c)/2$ is a constant.
Thus, Condition~\ref{con:ucc} holds with probability tending to one if $p \gg d\log d$ as $p\to\infty$.
\end{proposition}

\section{Statistical Recovery via Pairwise Dependence} \label{sec:stats}

We now consider the statistical problem of obtaining the parent-sharing graph $E_0$ and the model structure $A_0$ from a sample of observed data.
We utilize the heuristic that two observed variables that share a parent are likely to display a higher degree of association than observed variables that do not.
This motivates the use of pairwise dependence measures to determine if an edge in the parent-sharing graph should be present or not.

\subsection{Thresholdability}

Consider an arbitrary pairwise dependence measure $\rho(X_i, X_j)$, for which we will shorthand $\rho_{ij}$.
A simple criterion for recovering $E_0$ is if $\rho_{ij}$ determines membership of $(i, j) \in E_0$ by magnitude.
This idea can be formalized with the following definition.
\begin{definition}[Thresholdable]
We say $E_0$ is \emph{thresholdable} by a pairwise dependence measure $\rho(X_i, X_j)$ if there exists a $\tau_0 > 0$ such that
\begin{equation} \label{eq:separation}
\max_{(i,j)\notin E_0} \lvert \rho_{ij} \rvert \;<\; \tau_0 \;<\;
\min_{(i,j)\in E_0} \lvert \rho_{ij} \rvert.
\end{equation}
Equivalently, we may define a separation margin
\begin{equation} \label{eq:gap}
\gamma \coloneqq \dfrac{1}{2} \left[ \underset{(i,j) \in E_0}{\min}\lvert \rho_{ij} \rvert  - \underset{(i,j) \in E_0^c}{\max} \lvert \rho_{ij} \rvert \right],
\end{equation}
where $E_0$ is thresholdable by $\rho$ if and only if $\gamma > 0$.
\end{definition}

Then given a candidate threshold $\tau$, we can consider a sample estimate of $E_0$ as follows.
Let $r_{ij}$ be a sample estimate of $\rho_{ij}$.
Then
\begin{equation} \label{eq:E_hat}
\hat{E}(\tau) \coloneqq \{(i, j) : \lvert r_{ij} \rvert > \tau\}.
\end{equation}

We make a few remarks regarding specific choices of $\rho$.
In linear latent factor models, the Pearson correlation was studied in \cite{Kim2023}, where 
it is shown that $E_0$ is thresholdable by the Pearson correlation if the latent factors are independent. 
Subsequent work showed that thresholdability can be violated to a fairly large degree and the structure could still be recovered with a high amount of accuracy \citep{Kim2025}.

It is worth momentarily discussing thresholdability under the special case of independent latent factors.
If the factors are nonlinear, but independent, then any dependence measure that vanishes under independence can serve as a suitable $\rho$.
Such measures include distance correlation \citep{Szekely2009}, kernel-based dependence measures \citep{Gretton2005, Gretton2007}, or maximal correlation \citep{Renyi1959, Breiman1985}.
We state this in the following remark.
\begin{remark} \label{rem:indep_Z}
Suppose the latent variables $Z_1, \dots, Z_d$ are mutually independent.
If the dependence measure $\rho$ satisfies
\begin{equation}
\rho(X_i, X_j) = 0 \text{ if and only if } X_i \indep X_j,
\end{equation}
and if shared latent parents induce dependence among children, then $E_0$ is thresholdable by $\rho$.
\end{remark}

\subsection{Error Bounds and Consistency}

In this section, we derive the error bounds and consistency of the estimator $\hat{E}(\tau)$ and its implied structural estimator $\hat{A} = \psi(\hat{E}(\tau))$.
Suppose that $E_0$ is thresholdable by $\rho$ with margin $\gamma > 0$.
The key observation is that if the deviation error of $r_{ij}$ uniformly satisfies
\begin{equation}
\underset{(i, j)}{\max} \lvert r_{ij} - \rho_{ij} \rvert < \gamma,
\end{equation}
then the separation between edge and non-edge pairs is preserved.
Then there exists a threshold $\tau$ such that the estimator yields $\hat{E}(\tau) = E_0$.
The probability of recovery can then be controlled via a union bound over all pairs.
This is described in the following lemma.
\begin{lemma}[Error Bound for Edge Recovery] \label{lem:gen_bound}
If $E_0$ is thresholdable by $\rho$ and the deviation error $\lvert r_{ij} - \rho_{ij} \rvert$ has a concentration inequality of the form
\begin{equation}\label{eq:conie}
\Prob(\lvert r_{ij} - \rho_{ij} \rvert > \epsilon) \leq b(n, \epsilon),
\end{equation}
then the probability of error in edge set recovery has the bound
\begin{equation}
\Prob(\hat{E}(\tau_0) \neq E_0) \leq \binom{p}{2}b(n, \gamma).
\end{equation}
\end{lemma}

This general form of the bound allows us to easily study consistency.
First, under fixed $p$, consistency follows immediately if $b(n, \gamma) \rightarrow 0$ as $n \rightarrow \infty$, since the error probability vanishes.
More generally, if the number of observed variables is allowed to depend on the sample size, $p = p_n$, then we may use the following theorem.
\begin{theorem}[High-Dimensional Consistency] \label{thm:consistency}
Assume that $E_0$ is thresholdable by $\rho$ and the deviation error of $r_{ij}$ satisfies the concentration inequality~\eqref{eq:conie}  for all $(i,j)$. If $p_n = o(b(n, \gamma)^{-1/2})$, then we have
\begin{equation}
\lim_{n \to \infty} \Prob(\hat{E}(\tau_0) = E_0) = 1.
\end{equation}
If additionally the unique child condition  (Condition~\ref{con:ucc}) holds, then
\begin{equation}
\lim_{n \to \infty} \Prob(\hat{A} = A_0) = 1,
\end{equation}
where $\hat{A} = \psi(\hat{E}(\tau_0))$.
\end{theorem}

It is seen that consistency in recovering $E_0$ depends on the separation margin $\gamma$, the concentration behavior of the statistic $r_{ij}$, and the growth rate of $p_n$.
Essentially, a larger margin $\gamma$ allows for more rapid growth in $p_n$ while still ensuring consistency.
Consistency of $\hat{A}$ follows from the correspondence between the $\ch(Z_j)$ sets induced by ${A_0}$ in Equation~\ref{eq:pach} and the independent maximal cliques of $G(X, {E_0})$, which was established by Lemma~\ref{lem:ucc_imc_bijection}.

To provide a concrete example, we study Kendall's rank correlation coefficient \citep{Kendall1938}, which is useful if the components of $g(\cdot)$ are monotonic and the marginals of $X$ are continuous.
The estimator of this correlation coefficient is a bounded $U$-statistic \citep{Hoeffding1992}, therefore admitting a Hoeffding-type bound for $b(\cdot)$.
This can be used to obtain the following probability error bound.
\begin{example} \label{ex:hoeffding}
Assume that $X$ has continuous marginal distributions.
Take $\rho$ to be Kendall's rank correlation coefficient, and let $E_0$ be thresholdable by $\rho$.
Then
\begin{equation}
\Prob(\hat{E}(\tau_0) \neq E_0) \leq \binom{p}{2}2\exp\left( -\dfrac{ n \gamma^2}{8}\right).
\end{equation}
As long as $p=p_n = o(\exp(n\gamma^2 / 16))$, consistency is achieved.
\end{example}

\section{The Dependence Thresholding Algorithm} \label{sec:algorithm}

In Section~\ref{sec:structure_identifiability} we established that the latent factor structure can be recovered from a parent-sharing graph $E_0$, and in Section~\ref{sec:stats} we showed that this graph can be consistently estimated from pairwise dependence measures, given a suitable threshold $\tau$.
We turn our attention to the problem of choosing $\tau$, and consolidate the previous results into a structure learning algorithm given observed data.

The premise of the algorithm is a search over a set of candidate thresholds $\tau_k \in [0, 1]$ and analyzing each resulting estimated graph $G(X, \hat{E}(\tau_k))$ for independent maximal cliques.
Then using Lemma~\ref{lem:ucc_imc_bijection}, we estimate all the child sets $\ch(Z_j)$ to construct an estimate of the Jacobian support, $\hat{A}_k$.
A flexible nonlinear model is then fit, such as a neural network architecture constrained by $\hat{A}_k$, to obtain an estimate of the nonlinear mapping $\hat{g}_k$.
Then for each candidate model $\hat{g}_k$, we compute a model selection statistic, such as a loss from a validation set, which is used to select a final model and structure.
This procedure is formally described in the Dependence Thresholding (DT) Algorithm detailed in Algorithm~\ref{alg:correlation-thresholding}.

\begin{algorithm}
\caption{Dependence Thresholding Algorithm}\label{alg:correlation-thresholding}
\SetKwInOut{Input}{Inputs}
\SetKwInOut{Output}{Output}
\SetKwFor{For}{for}{}{endfor}
\SetKwIF{If}{ElseIf}{Else}{if}{}{else if}{else}{endif}
\Input{Sample pairwise dependence matrix $R$ and set of thresholds $\tau=\{\tau_k : k\in[m]\}$}
\Output{Jacobian support estimate $\hat{A}$ and mapping function estimate $\hat{g}$}
\For{$k\in[m]$}{
Calculate $\mathcal{G}(X,\hat{E}(\tau_k))$ and extract the set of independent maximal cliques $\mathcal{C}_k=\{C_1,\ldots,C_{|\mathcal{C}_k|}\}$\;
\label{step:extrac_imc}
Set $\hat{d}_k = |\mathcal{C}_k|$\;
\label{step:start_supp}
Initialize $\hat{A}_k = \varnothing$\;
\For{$(i,j)\in[p]\times[\hat{d}_k]$}{
\If{$i\in C_j$}{
add $(i,j)$ to $\hat{A}_k$\;
}
}
\label{step:end_supp}
Estimate a nonlinear mapping $\hat{g}_k$ subject to $\hat{d}_k$ latent variables and constrained by $\hat{A}_k$ (e.g., a neural network with constrained architecture)\;
\label{step:estimate}
}
Select one of the $m$ estimates from $\{\hat{g}_k : k\in[m]\}$ (e.g., using validation loss)\;
\label{step:model_select}
\end{algorithm}

\subsection{Structure Learning}

We make a few comments on implementation.
The extraction of an independent maximal clique in Step~\ref{step:extrac_imc} can be done by simply checking if the neighborhood of $X_i$ forms a clique.
If it does, then it forms an independent maximal clique with $X_i$ as a unique member \citep{Kim2023}.
This means the computational cost of this step is $O(k^2p)$, where $k$ is the maximum neighborhood size over all $X_i$.
This is in contrast with algorithms that enumerate all (non-independent) maximal cliques which have exponential complexity \citep{Eppstein2010}.

Note that Step~\ref{step:start_supp} to Step~\ref{step:end_supp} is an algorithmic implementation of the function $\psi(\cdot)$ to construct the estimated support from the graph.
It utilizes the correspondence between independent maximal cliques and children sets of the latent factors described in Lemma~\ref{lem:ucc_imc_bijection}.
This correspondence implies that the number of latent factors is simply the number of independent maximal cliques, and the members of those cliques are designated as the children of those latent variables.

\subsection{Estimation and Model Selection}

After a support is learned, this is passed to an estimation method in Step~\ref{step:estimate}, then a final model is chosen at Step~\ref{step:model_select} via a model selection procedure.
The methods of these two steps should be chosen in a compatible manner.
For example, likelihood-based model selection criteria (e.g., BIC) should be paired with maximum likelihood estimation methods.
If the estimation and model selection pair of methods are consistent, then the resulting model selected by the DT algorithm will also be consistent, assuming a valid $\tau$ is contained in the candidate thresholds.
Alternatively, the DT algorithm can be used by pairing flexible estimators, such as neural networks, and out-of-sample validation loss as model selection.
In this case, the procedure selects the candidate structure with the best out-of-sample performance.

We note that model estimation should be done in a manner that enforces the estimated Jacobian support $\hat{A}_k$.
For linear factor models, this is as straightforward as constraining the corresponding coefficient to zero.
For more general flexible methods, specifying proper architectures and subsequently estimating models with latent variables can be more complex.

A simple method for enforcing $\hat{A}_k$ can be by parameterizing each observed variable as a separate function of its admissible latent parents.
This technique was used by \cite{Ainsworth2018} to force certain latent variables to only affect certain groups of observed variables.
Alternatively, one may use a shared neural network architecture with an initial masking layer to restrict which latent variables affect which observed variables \citep{Moran2022}.
Conducting forward passes one-at-a-time per observed variable, with the correct mask active, ensures that the Jacobian support is enforced.

The remaining challenge is estimating the latent variables and model parameters.
This can be done either by incorporating stochastic nodes within a computational graph/network \citep[e.g., variational autoencoders;][]{Kingma2013, Rezende2014} or via expectation-maximization (EM) type procedures \citep{Dempster1977} that alternate between estimating latent variables and updating parameters.
Both approaches are compatible with the framework provided the estimated support is enforced.

For flexible estimators such as neural networks, model selection is typically performed using out-of-sample criteria.
In our implementation, we used validation log-likelihood to select among candidate structures.
More generally, cross-validation based losses (e.g., negative log-likelihood or mean squared error) may be used, which provide a practical means of selecting models with strong predictive performance within the candidate set.

\section{Simulation Studies} \label{sec:simulation}

We evaluated the DT algorithm in two simulations that assess (1) baseline performance under a neural network setting and (2) robustness to assumption violations.
In the first, we implemented the DT algorithm using a neural network for estimation and tested the performance of a validation-based model selection criterion.
In the second simulation, we evaluated the DT algorithm's ability to generate accurate candidate structures in high-dimensional settings, while violating thresholdability and the unique child condition.

\subsection{Neural Network Performance Simulation} \label{sec:nn_sim}

We generated data from a nonlinear latent factor model with additive errors:
\begin{equation}
X = g(Z) + \varepsilon.
\end{equation}
The latent variables followed a mean-zero, unit-variance, multivariate Gaussian distribution, with covariances defined by three directed acyclic graphs.
These graphs were a V-structure (common consequence; $d=3$) model, a chained mediation model ($d=4$), and a double mediation model ($d=4$).
These are illustrated in Figure~\ref{fig:sim_dgp}.
All path coefficients were set to 0.5.

The observed variables followed an independent-cluster structure (one parent per observed variable), with each latent factor having five observed variables as children.
Each mapping function was generated by cycling through a list of nonlinear functions, which is included in Appendix~\ref{app:mapping_fns}.
The noise variable $\varepsilon$ was Gaussian, with variance chosen such that a signal-to-noise ratio of $4:1$ would be achieved.
We generated 100 datasets per structure, using a total sample size of $n = 5000$, with $n_{\text{train}} = 4000$ used for model training and $n_{\text{valid}} = 1000$ set aside for model evaluation.

The DT algorithm used the Spearman rank correlation as the pairwise dependence measure.
A grid of 50 uniformly spaced values in the interval $[0, 1]$ was used as the thresholds for $\tau$.
Two selection strategies were compared over the candidate structures produced by the algorithm.
The first method was an oracle selector, where the most accurate structure on the solution path was selected.
The second was a validation likelihood selector.
After estimating the model using each candidate structure, we calculated the observed data log-likelihood from the validation set of data.
The structure with the highest validation log-likelihood was chosen as the estimated structure (see Appendix~\ref{app:validation} for details).

We used a stochastic EM method \citep{Celeux1985} for estimating the model.
We used the expected complete-data log-likelihood as the EM surrogate objective:
\begin{equation}
\begin{aligned}
Q(\theta \mid \theta^{(t)}) &= \E[\log f_{X, Z}(X, Z; \theta) \mid X; \theta^{(t)}]\\
&= \E[\log f_{X \mid Z}(X \mid Z; g, \Omega) + \log f_{Z}(Z; \Phi) \mid X; \theta^{(t)}],
\end{aligned}
\end{equation}
where $\theta = \{g, \Omega, \Phi\}$.
We used a neural network to estimate and represent $g$, which, together with $\Omega$, are the parameters for $f_{X \mid Z}$.
The parameters for $f_{Z}$ are contained in $\Phi$.
Our algorithm alternated between a stochastic E-step that consisted of a draw from the posterior $f_{Z \mid X}(Z \mid X; \theta^{(t)})$, and a partial M-step that used this draw to train the neural network $g$ and estimate $\Omega$.
We used a Metropolis-adjusted Langevin algorithm \citep[MALA;][]{Roberts1996} to obtain samples from the posterior.
We utilized a mask layer architecture \citep{Moran2022} to enforce the support estimates $\hat{A}_k$.
Further architecture details can be found in Appendix~\ref{app:architecture}.

After the selection of the structure, we collected the estimated number of latent variables $\hat{d}$, the $F_1$ accuracy of the estimated support $\hat{A}$, and the Hamming distance (HD) of $\hat{A}$.
Our $F_1$ and HD calculations took into account that latent factor labels can be arbitrarily permuted, and these details are included in Appendix~\ref{app:outcomes}.
The results are displayed in Table~\ref{tab:results}.
Across all three data-generation settings, the DT algorithm achieved near-perfect structure recovery under oracle selection.
Mean $F_1$ scores ranged from 0.988 to 0.994, while HD remained below 0.5 in all settings.
Estimated latent dimensionality was also highly accurate, with mean $\hat{d}$ within 0.07 of the truth for all three conditions.
The validation likelihood for model selection led to slightly lower but competitive performance.
Mean $F_1$ scores ranged from 0.952 to 0.964, with HD between 1.4 and 2.13 out of 45 possible entries.
Estimated dimensionality remained close to the true value, with small upward biases of at most 0.3.

\begin{table}
\caption{Results of the simulation experiment under two structure selection mechanisms and three data-generating processes of $Z$ (denoted under Model).
The average of 100 datasets is displayed alongside one standard error for the estimated number of latent factors, the $F_1$ accuracy score, and HD.}
\label{tab:results}
\centering
\begin{tabular}{llcccc}
\toprule
Selection & Model & $d$ & $\hat d$ & $F_1$ & HD \\
\midrule
\multirow{3}{*}{Oracle}
  & V-Structure & 3 & $3.03 \pm 0.018$ & $0.993 \pm 0.004$ & $0.233 \pm 0.090$ \\
  & Chained Mediation & 4 & $4.07 \pm 0.037$ & $0.994 \pm 0.003$ & $0.267 \pm 0.146$ \\
  & Double Mediation & 4 & $4.03 \pm 0.018$ & $0.988 \pm 0.005$ & $0.467 \pm 0.185$ \\
\midrule
\multirow{3}{*}{Likelihood}
  & V-Structure & 3 & $3.13 \pm 0.346$ & $0.958 \pm 0.061$ & $1.400 \pm 2.11$ \\
  & Chained Mediation & 4 & $4.20 \pm 0.484$ & $0.964 \pm 0.061$ & $1.600 \pm 2.81$ \\
  & Double Mediation & 4 & $4.30 \pm 0.651$ & $0.952 \pm 0.066$ & $2.130 \pm 3.05$ \\
\bottomrule
\end{tabular}
\end{table}

These results demonstrate that the DT algorithm can be effectively used with flexible nonlinear estimators to recover latent structure in practice.
The validation-based selection criteria displayed high accuracy, achieving fairly comparable performance to the oracle selector.
This suggests that the thresholding procedure generates candidate structures that are sufficiently close to the truth for standard validation criteria to recover the correct model.

\subsection{High-Dimensional Recovery and Robustness}

We next examine the robustness of the DT algorithm to violations of the thresholdability and unique child condition (UCC) assumptions in high-dimensional settings.
We generated data under the same nonlinear factor model as in the previous simulation, but used smaller sample sizes $n \in \{250,500,1000\}$, while scaling $p=1.5n$ and $d=0.1n$.

To study the effect of violating thresholdability, we varied the off-diagonal entries of the latent correlation matrix across the values $\phi_{ij} \in \{0, 0.25, 0.75\}$.
As latent factors become more correlated, variables that do not share parents become dependent through correlated parents, shrinking the thresholdability separation margin.
To study the effect of violating the UCC, we constructed models in which a proportion of latent variables lacked unique children.
This was done by starting from an independent cluster structure, and designating a fraction $\{0,0.2,0.4\}$ of latent variables to violate the UCC by randomly assigning its children an additional parent.
For this simulation, we only used the oracle selection method to isolate the effect of assumption violations, and for computational feasibility.

\begin{figure}
\centering
\includegraphics[width=\textwidth, keepaspectratio]{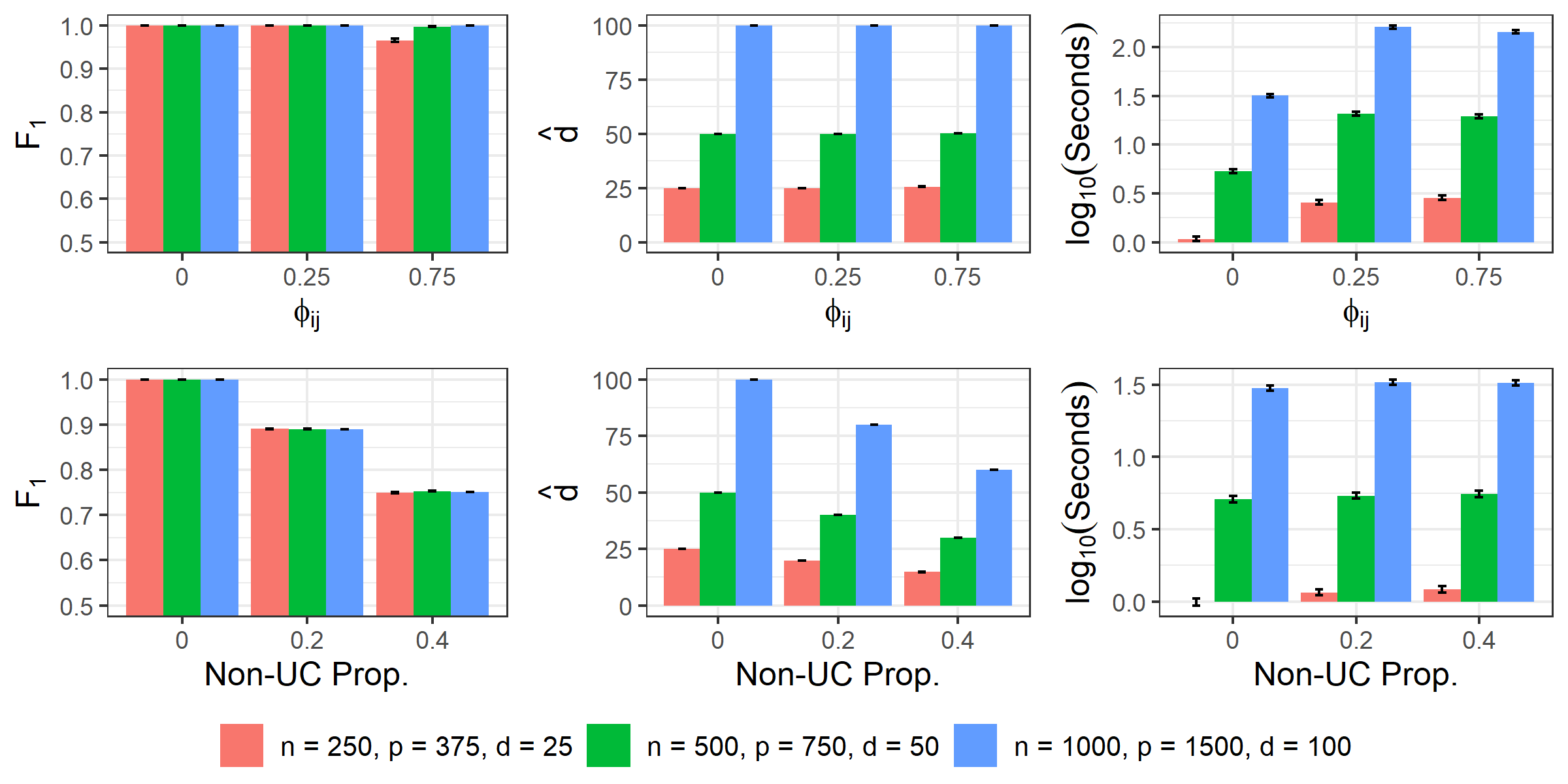}
\caption{High-dimensional simulation results showing $F_1$ accuracy, the number of recovered latent factors $\hat{d}$, and the computation time in $\log_{10}$ seconds.
Note that the $F_1$ accuracy display ranges from 0.5 to 1.0 to better show variation.
The top row depicts algorithm performance across levels of $\phi_{ij}$.
The bottom row depicts performance across proportions of latent variables with non-unique children.
Black bars indicate $\pm 1$ standard error.}
\label{fig:high_dim}
\end{figure}

We collected the $F_1$ accuracy, $\hat{d}$, and computation time across 100 datasets per condition.
These results are displayed in Figure~\ref{fig:high_dim}.
Under conditions where we varied the latent correlations, the method achieves near-perfect recovery across all model sizes and $\phi_{ij}$ values, as well as accurate estimation of the number of latent factors.
Only a minor degradation of performance was observed in the $\phi_{ij}=0.75$ setting for the smallest sample size ($n=250$).
For violations of the UCC, performance degraded as the proportion of non-unique-child factors increased.
Under mild violations (non-UCC $= 0.2$), the method still recovered most of the latent structure, with average $F_1$ scores of approximately $0.9$, though the number of latent factors was underestimated (recovering about $80\%$ of the true number on average).
Under more substantial violations (non-UCC $= 0.4$), performance declined more noticeably, with $F_1 \approx 0.75$ and roughly $60\%$ of the latent factors recovered on average.
Unsurprisingly, the computation time scaled with model size, but remained fairly reasonable where the worst case average was less than three minutes.

These results indicate that the DT algorithm is robust to thresholdability violations.
For the UCC, mild departures leave most of the structure intact, whereas more substantial violations lead to clearer degradation in both $F_1$ and the estimated number of latent factors.
These results indicate that while UCC remains important for exact recovery, partial recovery remains possible even when the assumption is violated.

\section{Discussion} \label{sec:discussion}

In this work, we studied a graph-theoretic method for identifying and learning the structure of a general latent factor model from observational data alone.
We established a correspondence between the latent structure and a parent-sharing graph representation of the model, which can be recovered through pairwise dependence measures.
Further, we derived high-dimensional recovery error bounds and high-dimensional conditions for consistency.
Putting these ideas together, we proposed the DT algorithm as a practical method for learning the latent factor structure in practice.
Our simulation studies demonstrated near-perfect performance when the assumptions are met.
When they are not, the method remained robust to thresholdability violations, while violations of the UCC led to gradual degradation, with meaningful partial recovery under mild departures and more substantial losses under stronger ones.

Several limitations remain, namely the assumptions of thresholdability and the unique child condition.
While our simulations showed that the DT algorithm can tolerate mild to moderate violations of these assumptions, further work is required to relax these settings, and possibly to obtain guarantees for partial recovery.

\bibliographystyle{asa}
\bibliography{references}

\newpage
\appendix
\appendixpage

\section{Proofs}

\subsection{Proof of Lemma~\ref{lem:ucc_imc_bijection}} \label{app:ucc_imc_bijection_proof}

Aside from the modified definition of $\mathrm{ch}(Z_j)$ used in the present setting, the proof is identical to that of Lemma~4.3 of \citep{Kim2023}.

\subsection{Proof of Proposition~\ref{prop:ucc_random}} \label{app:ucc_random_proof}
Fix $j \in [d]$ and $i \in [p]$.
If $X_i$ has at least one parent, then
\begin{equation}
\Prob(X_i \text{ is a unique child of } Z_j) = \frac{\alpha(1-\alpha)^{d-1}}{1-(1-\alpha)^{d}} \geq \alpha(1-\alpha)^{d}.
\end{equation}
If edge probabilities are independent across observed variables then
\begin{equation}
\Prob(Z_j \text{ has no unique child}) \leq \bigl[1 - \alpha(1-\alpha)^{d}\bigr]^{p} \leq \exp\bigl(-p\alpha(1-\alpha)^{d}\bigr),
\end{equation}
where the last step uses $1-x \le e^{-x}$.
A union bound over the $d$ latent factors gives
\begin{equation}
\Prob(\mathrm{UCC}) \geq 1 - d\exp\bigl(-p\alpha(1-\alpha)^{d}\bigr).
\end{equation}

For the sparse regime $\alpha = c/d$, with a fixed $c > 0$, we have
\begin{equation}
\alpha(1-\alpha)^{d} = \dfrac{c}{d} \left(1-\dfrac{c}{d}\right)^{d}
 \geq \dfrac{c\exp(-c)}{2d} = \dfrac{c_1}{d},
\end{equation}
for sufficiently large $d$ and letting $c_1 = c \exp(-c)/2$.
This yields
\begin{equation}
\Prob(\mathrm{UCC}) \geq 1 - d\exp\left(-c_1 p / d\right).
\end{equation}
If $p \gg d\log d$, then $d\exp\left(-c_1 p / d\right) \to 0$ as $p \to \infty$ and Condition~\ref{con:ucc} holds with probability tending to one.

\subsection{Proof of Remark~\ref{rem:indep_Z}}
If $(i,j) \notin E_0$, then $X_i$ and $X_j$ do not share latent parents.
Since the latent variables and errors are mutually independent, this implies $X_i \indep X_j$, and hence $\rho(X_i,X_j)=0$.

If $(i,j)\in E_0$, then $X_i$ and $X_j$ share at least one latent parent, and we assume sharing latent parents implies $X_i\not\!\indep X_j$.
Since $\rho(X_i,X_j)=0$ if and only if $X_i\perp X_j$, we have $\rho(X_i,X_j)\neq 0$.

Thus,
\[
\max_{(i,j)\in E_0^c}|\rho(X_i,X_j)|=0
<
\min_{(i,j)\in E_0}|\rho(X_i,X_j)|.
\]
Therefore, there exists a threshold $\tau_0$ between these two quantities, so $E_0$ is thresholdable by $\rho$.

\subsection{Proof of Lemma~\ref{lem:gen_bound}} \label{app:gen_bound_proof}
We assume $E_0$ is thresholdable by $\rho$, and thus we have a margin of separation $\gamma > 0$.
If the deviation error of $r_{ij}$ has the form
\[
\Prob(\lvert r_{ij} - \rho_{ij} \rvert > \epsilon) \leq b(n, \epsilon),
\]
then by the union bound we have
\[
\begin{aligned}
\Prob(\underset{(i, j)}{\max} \lvert r_{ij} - \rho_{ij} \rvert > \epsilon) &\leq \binom{p}{2}b(n, \epsilon)\\
\Rightarrow \Prob(\underset{(i, j)}{\max} \lvert r_{ij} - \rho_{ij} \rvert > \gamma) &\leq \binom{p}{2}b(n, \gamma)\\
\Rightarrow \Prob(\hat{E}(\tau_0) \neq E_0) &\leq \binom{p}{2}b(n, \gamma).
\end{aligned}
\]

\subsection{Proof of Theorem~\ref{thm:consistency}} \label{app:consistency_proof}
Let $p_n$ depend on $n$.
By Lemma~\ref{lem:gen_bound}, we have 
\[
\Prob(\hat{E}(\tau_0) \neq E_0) \leq \binom{p_n}{2}b(n, \gamma).
\]
This implies that
\[
\lim_{n \to \infty} \Prob(\hat{E}(\tau_0) = E_0) = 1
\]
if
\[
\lim_{n \to \infty} \binom{p_n}{2}b(n, \gamma) = 0.
\]
We first observe that
\[
p_n = o(b(n, \gamma)^{-1/2}) \Rightarrow p^2_n b(n, \gamma) \rightarrow 0,
\]
as $n \rightarrow \infty$.
Then we have
\[
\binom{p_n}{2} b(n, \gamma) \leq \dfrac{p_n^2}{2} b(n, \gamma) \leq p_n^2 b(n, \gamma) \rightarrow 0
\]
as $n \rightarrow \infty$.
Thus if $p_n = o(b(n, \gamma)^{-1/2})$, we have consistency of $\hat{E}(\tau_0)$ under the conditions of Lemma~\ref{lem:gen_bound}.

Under the unique child condition, Lemma~\ref{lem:ucc_imc_bijection} gives a map $\psi$ such that $\psi(E_0) = A_0$, thus yielding $\lim_{n \to \infty} \Prob(\hat{A} = A_0) = 1$.

\subsection{Proof of Example~\ref{ex:hoeffding}} \label{app:hoeffding_proof}

We begin with the Hoeffding-type concentration inequality for $U$-statistics \citep{Serfling1980}
\[
\Prob(\lvert U - \E[U] \rvert \geq t) \leq 2\exp\left( -\dfrac{2 \lfloor n/m \rfloor t^2}{(b - a)^2}\right),
\]
where $m$ is the order of the $U$-statistic, and $a \leq h(x_1, \dots, x_m) \leq b$ are the bounds on the kernel function $h$.
In the case of Kendall's rank correlation coefficient, we have $m = 2$, with $a = -1$ and $b = 1$ \citep{Hoeffding1992}.
Thus we have
\[
\begin{aligned}
\Prob(\lvert r_{ij} - \rho_{ij} \rvert \geq t) &\leq 2\exp\left( -\dfrac{ \lfloor n/2 \rfloor t^2}{2}\right)\\
&\leq 2\exp\left( -\dfrac{ n t^2}{8}\right),
\end{aligned}
\]
since $\lfloor n/2 \rfloor \geq n/4$ for $n \geq 2$.
Taking this quantity as $b(n, \gamma)$ where $\gamma = t$, we have from Lemma~\ref{lem:gen_bound}
\[
\Prob(\hat{E}(\tau_0) \neq E_0) \leq \binom{p}{2}2\exp\left( -\dfrac{ n \gamma^2}{8}\right).
\]
Then if we allow $p$ to grow with $n$, following Theorem~\ref{thm:consistency}, consistency is maintained if
\[
\begin{aligned}
p_n &= o(b(n, \gamma)^{-1/2})\\  
&= o(\exp( -n \gamma^2 / 8)^{-1/2})\\
&= o(\exp(n \gamma^2/16))
\end{aligned}
\]

\section{Additional Simulation Details}

\begin{figure}
\centering
\includegraphics[width=\textwidth, keepaspectratio]{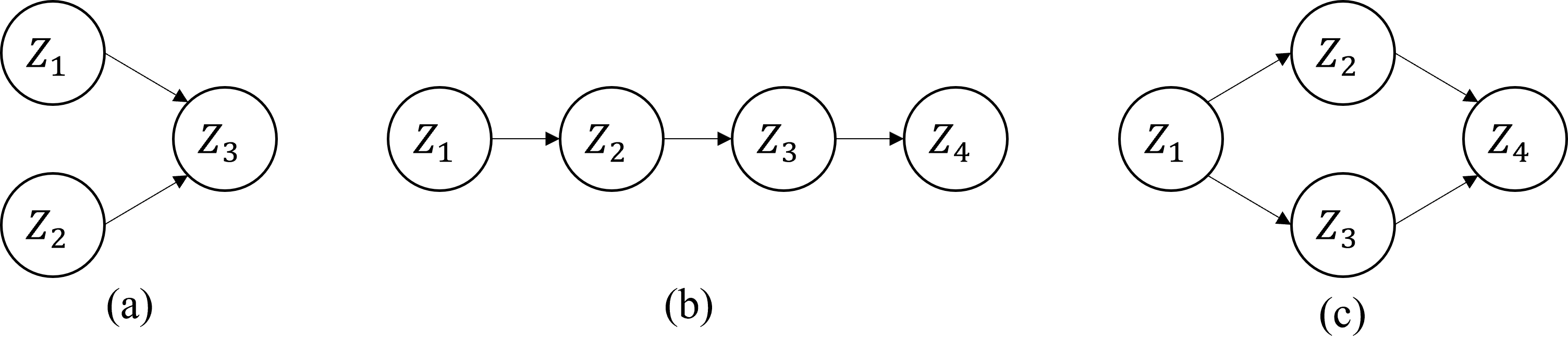}
\caption{Data-generating processes for latent factors.
The processes were (a) common consequence (V-structure), (b) chained mediation, and (c) double mediation.
All path coefficients were set to 0.5.}
\label{fig:sim_dgp}
\end{figure}

Illustrations of the latent variable generating process for Section~\ref{sec:nn_sim} are depicted in Figure~\ref{fig:sim_dgp}.

\subsection{Mapping Functions} \label{app:mapping_fns}

To generate each observed variable from a given latent variable $Z_j$, we cycled through the following nonlinear functions.

\[
\begin{aligned}
g_1(Z_j) &= 1.5 \, Z_j + \sin(0.5 \, Z_j) & g_6(Z_j) &= \sinh^{-1}(Z_j)\\
g_2(Z_j) &= \operatorname{erf}(Z_j) & g_7(Z_j) &= \tan^{-1}(Z_j)\\
g_3(Z_j) &= Z_j(1 + |Z_j|)^{-1} & g_8(Z_j) &= \exp\left(0.5 \, Z_j\right)\\
g_4(Z_j) &= \log\left( \exp(Z_j) + 1 \right) & g_9(Z_j) &= \tanh\left(0.8 \, Z_j\right) \\
g_5(Z_j) &= \operatorname{sign}(Z_j) \sqrt{|Z_j| + 10^{-3}} & g_{10}(Z_j) &= (1 + \exp(Z_j))^{-1}
\end{aligned}
\]

In the case of multiple parents, each observed variable was constructed additively from nonlinear contributions from all of its parents.
\[
X_i = \sum_{j \in \pa(X_i)} g_k(Z_j) + \varepsilon_i,
\]
where the index $k$ cycles through $\{1, \dots, 10\}$ per use.

\subsection{Outcomes} \label{app:outcomes}

To compare the estimated and true Jacobian supports ($\hat{A}$ vs. $A_0$) we computed the minimum HD over all permutations of latent variable labels of $\hat{A}$.
That is, we define an HD as
\[
\mathrm{HD}(\widehat{A}, A_0) \coloneqq \min_{\pi}\left| \widehat{A}^{\pi} \triangle A_0 \right|
\]
where
\[
\widehat{A}^{\pi} \coloneqq \{(i,\pi(j)) : (i,j)\in \widehat{A}\},
\]
where $\triangle$ denotes the symmetric difference between two sets.
The permutation $\pi$ reconciles the fact that latent factor indices are arbitrary, and thus the ordering of the estimated latent variables need not coincide with the ordering of the true latent variables. Equivalently, $\mathrm{HD}(\widehat{A}, A_0)$ is the minimum number of support additions and deletions required to transform $\widehat{A}$ into $A_0$ over all permutations of the latent indices.

In addition to HD, we also report the $F_1$ score, a normed measure of classification.
This allows for comparability between models with differing dimensions of $A$, that is differing $p$ and $d$.
Note that the $F_1$ score is simply the harmonic mean between precision and recall.
Once again using permutation matrices to reconcile different orderings of $Z$, we have
\[
F_1(\widehat{A}) \coloneqq \max_{\pi} \left[ \dfrac{2|\widehat{A}^{\pi}\cap A_0|}{2|\widehat{A}^{\pi}\cap A_0| + |\widehat{A}^{\pi}\triangle A_0|}\right]
\in [0,1],
\]
where the maximum is taken over all permutations $\pi$ of the latent factor indices.

\section{Estimation}

\subsection{Neural Network Architecture} \label{app:architecture}

Our neural network architecture was a fully connected feedforward network, with three hidden layers of width 100, each followed by $\tanh()$ activations.
The network takes a $d$-dimensional input vector $z$ and yields a $p$-dimensional output vector $x$.
We used the masking technique of \citep{Moran2022} to enforce the Jacobian support, $A$.
For each observed variable $x_i$, define a mask variable
\[
w_i \coloneqq (w_{i1}, \dots, w_{id}) \in \{0, 1\}^d
\]
where
\[
w_{ij} = \begin{cases}
  1 & \text{if }(i, j) \in A\\
  0 & \text{otherwise.}
\end{cases}
\]
Then forward passes are performed separately per observed variable, where $z$ is first passed through the mask via elementwise multiplication
\[
\tilde{z}_i \coloneqq z \odot w_i,
\]
and the masked vector $\tilde{z}_i$ is used as the input layer to the shared network.
This ensures that the Jacobian support is enforced.

\subsection{Stochastic EM Optimization}

Network parameters were estimated using a stochastic expectation-maximization (EM) procedure.
In the stochastic E-step, latent variables were sampled from the posterior distribution $f_{Z \mid X}(Z \mid X; \theta^{(t)})$ using the Metropolis-adjusted Langevin algorithm (MALA), where $\theta = \{g, \Omega, \Phi\}$.
A single post burn-in sample was taken to create a completed dataset, then passed to the M-step.
In the M-step, neural network parameters were updated by maximizing complete-data log-likelihood using the Adam optimizer.
We conducted 3000 EM iterations per dataset.

For the MALA sampler, proposal step sizes were adaptively tuned during sampling to target an average acceptance rate of $0.574 \pm 0.01$, which is the value recommended by \citep{Roberts1998}.
Once achieved, the step size was fixed and the chain was advanced for an additional 10 iterations, which were discarded before drawing a final sample.
This was done to reduce the dependence between the adaptive mechanism and the retained sample.
After using this sample in the M-step, this state of the chain was used to warm-start the MALA sampler for the next iteration.

\subsection{Validation Model Selection} \label{app:validation}
Upon conclusion of estimation, a validation log-likelihood, $\log f(x_{\text{valid}})$ was calculated.
This was done using the reciprocal identity:
\[
\dfrac{1}{f(x)} = \mathbb{E}_{Z \mid X} \left[\frac{h(z)}{f(x,z)}\right],
\]
where $h(z)$ is an auxiliary density that should ideally be close to $f(Z \mid X)$.
This yields a Monte Carlo estimator per validation sample:
\[
\log \hat{f}(x_i) = -\log \left(\dfrac{1}{m}\sum_{k=1}^m \dfrac{h_i(z_i^{(k)})}{f(x_i,z_i^{(k)})}\right),
\]
over $m = 1000$ posterior samples.
The auxiliary density was a Gaussian density per observation as
\[
h_i(z) = \mathcal{N}(z; \hat\mu_i, \hat\Sigma_i),
\]
whose parameters were estimated from the set of MALA posterior draws for that observation as
\[
\begin{aligned}
\hat{\mu}_i &= \dfrac{1}{m}\sum_{k = 1}^m z_i^{(k)}\\
\hat{\Sigma}_i &= \dfrac{1}{m - 1}\sum_{k = 1}^m (z_i^{(k)} - \hat\mu_i)(z_i^{(k)} - \hat\mu_i)^T.
\end{aligned}
\]
Then summing over all validation samples we have
\[
\log \hat{f}(x_{\text{valid}}) = \sum_{i=1}^{n_{\text{valid}}} \log \hat{f}(x_i),
\]
for all samples $i \in [n_{\text{valid}}]$.

\end{document}